\documentclass[twocolumn]{aastex63}
\usepackage{amsmath}
\received{}
\revised{}
\accepted{}
\submitjournal{ApJ}

\shorttitle{Variability in the Great Disk Shadow}
\shortauthors{Pontoppidan et al.}

\begin{document}

\title{Variability of the Great Disk Shadow in Serpens}

\correspondingauthor{Klaus Pontoppidan}
\email{pontoppi@stsci.edu}

\author[0000-0002-0786-7307]{Klaus M. Pontoppidan}
\affil{Space Telescope Science Institute \\
3700 San Martin Drive \\
Baltimore, MD 21218, USA}

\author{Joel D. Green}
\affiliation{Space Telescope Science Institute \\
3700 San Martin Drive \\
Baltimore, MD 21218, USA}

\author{Tyler A. Pauly}
\affiliation{Space Telescope Science Institute \\
3700 San Martin Drive \\
Baltimore, MD 21218, USA}

\author{Colette Salyk}
\affiliation{Vassar College \\
124 Raymond Ave \\
Poughkeepsie, NY 12604, USA }

\author{Joseph DePasquale}
\affiliation{Space Telescope Science Institute \\
3700 San Martin Drive \\
Baltimore, MD 21218, USA}

\begin{abstract}

We present multi-epoch Hubble Space Telescope imaging of the Great Disk Shadow in the Serpens star-forming region. The near-infrared images show strong variability of the disk shadow, revealing dynamics of the inner disk on time scales of months. The Great Shadow is projected onto the Serpens reflection nebula by an unresolved protoplanetary disk surrounding the young intermediate-mass star SVS2/CK3/EC82. Since the shadow extends out to a distance of at least 17,000 au, corresponding to a light travel time of 0.24 years, the images may reveal detailed changes in the disk scale height and position angle on time scales as short as a day, corresponding to the angular resolution of the images, and up to the 1.11 year span between two observing epochs. We present a basic retrieval of temporal changes in the disk density structure, based on the images. We find that the inner disk changes position angle on time scales of months, and that the change is not axisymmetric, suggesting the presence of a non-axisymmetric dynamical forcing on $\sim$1\,au size scales. We consider two different scenarios, one in which a quadrupolar disk warp orbits the central star, and one in which an unequal-mass binary orbiting out of the disk plane displaces the photo-center relative to the shadowing disk. Continued space-based monitoring of the Serpens Disk Shadow is required to distinguish between these scenarios, and could provide unique, and detailed, insight into the dynamics of inner protoplanetary disks not available through other means.
\end{abstract}
  
\keywords{Exoplanet formation --- Protoplanetary Disks --- Time Domain Astronomy}

\section{Introduction} \label{sec:intro}

Pre-main sequence stars younger than 1-2 Myr often illuminate nearby dust from their natal molecular cloud \citep{Kenyon95}. This material could originate from a remnant envelope of the young star itself, or could be a quiescent part of the molecular cloud encountered by the young star after traveling a significant distance from its birthplace \citep[e.g.][]{Britt16}. Indeed, a young star may cross its parent cluster within $\sim$1 Myr \citep{Kraus08, Zari19}, which is significantly less than the lifetime of the parent molecular cloud. It is plausible that many young stars will illuminate quiescent parts of a cloud at some point in their lifetime. Pre-main sequence stars are typically surrounded by protoplanetary disks, and since such disks are generally highly optically thick at optical/NIR wavelengths, they may cast a shadow on the reflection nebula, especially if viewed close to edge-on \citep{Hodapp04,Pontoppidan05}. The projection of the disk onto a large reflection nebula can greatly magnify small structure in the obscuring disk. Indeed, the apparent angular size of such {\it disk shadows} is only limited by the size of the reflection nebula illuminated by the central star, and may be orders of magnitude larger than the protoplanetary disk itself. Disk shadows therefore present a unique opportunity to explore the geometry of disks on scales otherwise not resolved by direct imaging, primarily the disk scale height, inclination, and position angle. 

\begin{figure*}[ht!]
\centering
\includegraphics[width=16cm]{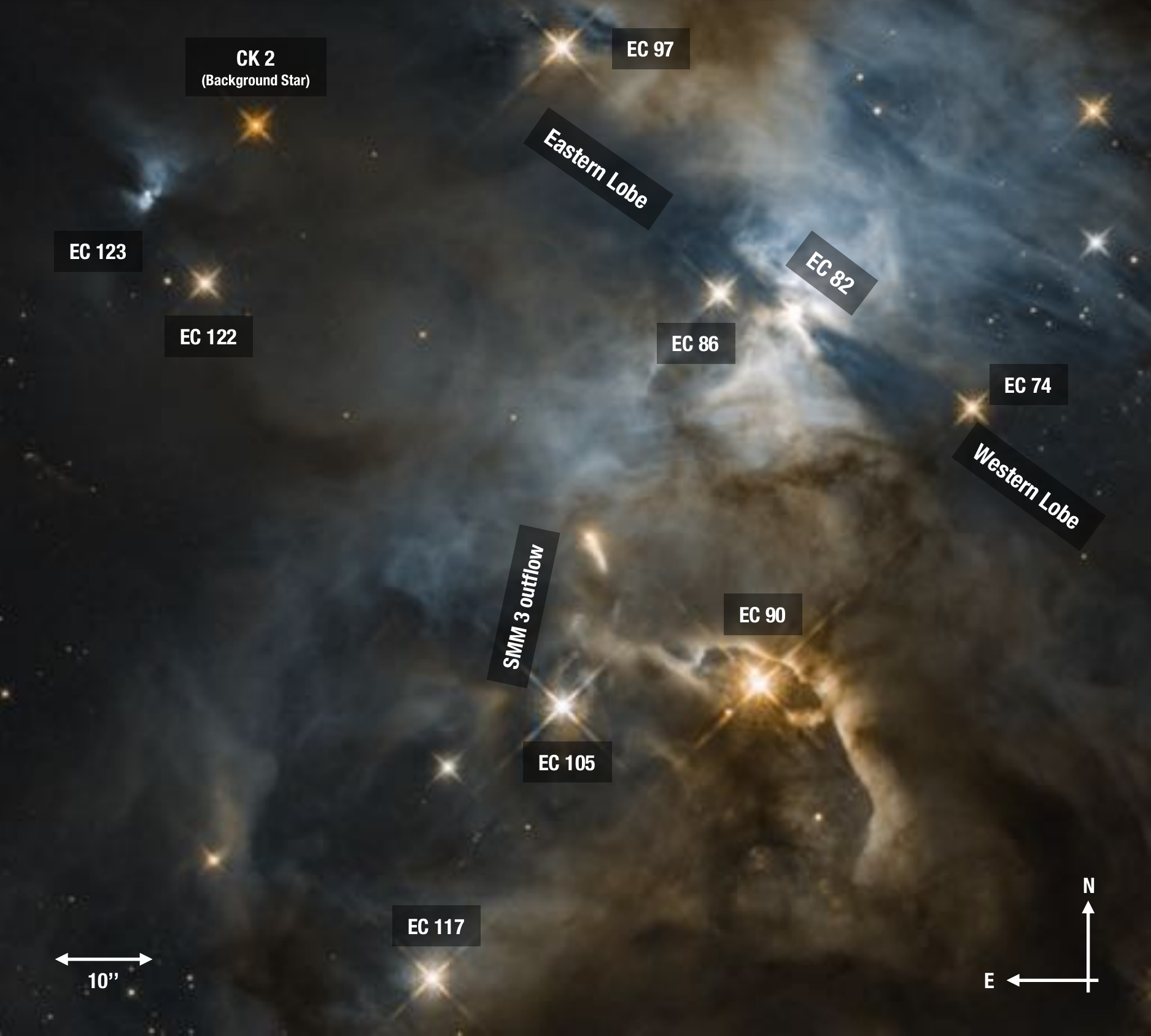}
\caption{Annotated two-color WFC3 image of the Serpens core in the vicinity of EC 82, in F125W (blue) and F164N (red). [A higher-resolution version of this figure is available in the journal article.]}
\label{fig:ec82-annotated}
\end{figure*}

One of the most iconic disk shadows is the great shadow in the main core of the Serpens star-forming region (Figure \ref{fig:ec82-annotated}), illuminated by the young intermediate-mass star EC 82 (HBC 672/CK 3/SVS 2) \citep{Pontoppidan05}. Indeed, EC 82 is the dominant illuminating source in the Serpens main core reflection nebulosity (SRN) \citep{Sugitani10}. Because the near-infrared spectrum of EC 82 is strongly veiled, its effective temperature is uncertain \citep[see discussion in][]{Gorlova10}. \cite{Winston09} report an effective temperature of 3900\,K (K8) with H- and K-band spectra, but both \cite{Doppmann05} and \cite{Gorlova10} find that the near-infrared spectrum of EC 82 is too veiled to yield a reliable effective temperature, and the featureless spectrum of EC 82 is therefore also consistent with an early-type photosphere. \cite{Pontoppidan05} used a detailed radiative transfer model of the shadow, spectral energy distribution, and reflection nebula to estimate a luminosity of 30\,$L_{\odot}$ (scaled to the VLBI distance of $436\pm 9$\,pc \citep{OrtizLeon17} from their assumed distance of 250\,pc). Using the pre-main sequence evolutionary track of \cite{Siess00} and assuming an age of 1-2 Myr predicts effective temperatures of 5400-6100\,K, and stellar masses of 2.5-3.0\,$M_{\odot}$. That is, based in the accurate VLBI distance, EC 82 is likely a young intermediate-mass star, and either an actual, or a precursor to, a Herbig Ae star.

In this paper we present near-infrared Hubble Space Telescope / Wide-Field Camera 3 (HST-WFC3) imaging of the Serpens disk shadow from EC 82 over two epochs separated by 404 days, and demonstrate that the shadow varies significantly over this timescale. We find that the angle of the shadow changed by several degrees, providing a unique probe of the dynamics of the inner disk of this system. We also find evidence for significant variability within at least one of the two epochs on time scales of $\sim$1 month. 

Time-variable self-shadowing of outer protoplanetary disks has been observed by direct imaging in a variety of systems. Such phenomena may related to that described here, but the order-of-magnitude magnification provided by the Serpens reflection nebula allows investigation of phenomena on spatial scales not possible in directly observed disks. For instance, the HH30 edge-on disk exhibited asymmetry and variability when observed by HST between 1994 and 2005 \citep{Watson07}, with possible (but poorly constrained) periodicity on timescales less than 1 yr. Using HST/STIS and NICMOS imaging, \citet{Debes17} found azimuthal asymmetry associated with $\sim$ 15.9 yr periodic variability in the more evolved TW Hya system between 1998 and 2005. In that case, they hypothesized that the disk interior to 1 au was inclined and precessing relative to the outer disk, causing a shadow on the outer disk. They invoke an external perturber as the cause of the disk warp or misalignment. Using VLT/SPHERE, \citet{Pinilla18} found aperiodic outer disk shadowing around the Dipper Star RXJ1604.3-2130, consistent with observed dimming events. In this case, either a planetary mass companion or magnetic field alignment effects could be invoked. In each case, the variability of the shadow over a large spatial scale (10s of au) represented a much smaller scale ($\sim$ 1 au) perturbation in the disk. 

In Section \ref{sec:obs}, we describe the two epochs of observations. In Section \ref{sec:analysis}, we quantify the shadow morphology and its variability using the high resolution and reproducibility of the HST data. Finally, in Section \ref{sec:discussion}, we discuss the implications for the EC 82 protoplanetary disk. 

\section{Observations}
\label{sec:obs}
EC 82 was first observed by WFC3 on 22 July 2017 as part of HST program 14181 using the F160W filter. The region was subsequently observed in the F125W and F164N filters on 30 August 2018 as part of program 15597. These observations covered a $123''\times 136''$ field of view, at a resolution of 0\farcs13 per pixel. For the F160W observation, the STEP50 sample sequence with 14 non-destructive detector reads was used, while the STEP400 sequence with 14 reads was used for the F164N narrow-band image. Table \ref{table:obslog} summarizes the observations. 

\begin{deluxetable}{lccc}[ht]
\tablecolumns{4}
\tablewidth{0pc}
\tablecaption{Observing log\label{table:obslog}}
\tablehead{
\colhead{Date} & \colhead{Instrument} & \colhead{Filter} & \colhead{Exposure time} \\
\colhead{}     & \colhead{}           & \colhead{}       & \colhead{[seconds]}
}
\startdata
2017 Jul 22  & WFC3 & F160W & 1597 \\
2018 Aug 30  & WFC3 & F164N & 5998 \\
2018 Aug 30  & WFC3 & F125W & 1798 \\
\enddata
\end{deluxetable}

Both data sets were downloaded from the Mikulski Archive for Space Telescopes (MAST) in FLT format, as output from the {\tt calwf3} data reduction pipeline, which includes basic calibration of the raw data, including bias and dark current subtraction, linearity correction, flat fielding, bad pixel masking, and cosmic ray removal. Each image was processed through the {\tt Drizzlepac} package, using {\tt Tweakreg} for individual dithered exposure alignment, and {\tt Astrodrizzle} to combine the individual exposures. The {\tt Astrodrizzle} processing included sky subtraction using a median statistic, along with further cosmic ray reduction using {\tt driz\_cr}, finally creating a drizzled, combined image with an output pixel size of 0\farcs08.

Figure \ref{fig:ec82-annotated} shows the two-color composite of the Serpens core (F125W, blue and F164N, red), with prominent sources indicated. Most bright stars are cluster members, except for CK 2, which is a well-known background red supergiant behind 46 magnitudes of visible extinction \citep{Casali96}. Also clearly visible is the outflow from the class 0 protostar SMM 3 (the source itself is not visible in the near-infrared). Dominating the reflection nebulosity is the  matter surrounding the class I binary EC 90 to the south and the Serpens Reflection Nebula to the north, which is transversed by the EC 82 disk shadow. The blue color of the EC 82 reflection nebula indicates that there is relatively little foreground extinction compared to the redder nebulosity around EC 90. It is curious that another likely disk shadow is visible to the south-east of CK 2. This shadow surrounds the low-luminosity young star EC 123, coincidentally with almost the same position angle as the EC 82 shadow. 

\section{Analysis}
\label{sec:analysis}
 The EC 82 disk shadow consists of two opposite lobes, one extending toward the north-east, and one extending to the south-west (henceforth referred to as the ``eastern'' and ``western''  lobes, respectively). Together, the shadow lobes have a position angle of $\sim50\degr$ east of north. The shadow can be traced to a distance of more than  17,000~au (40\arcsec) from the central source along each lobe, corresponding to a light-travel time of 0.27 years, or almost 100 days. Thus, a change in the geometry of the inner disk, or its central light source, on time scales longer than 100 days can lead to changes in the shadow across its full extent when comparing the two epochs. Conversely, changes in the disk taking place on time scales less than 100 days will manifest as changes over a smaller range of radii in the shadow, as the perturbation travels outwards at the speed of light. However, the shadow is best-defined within $\sim 45$~days and $\sim 75$~days in the western and eastern lobes, respectively, and we therefore restrict our quantitative analysis to 45 days. Given the angular resolution of HST at 1.6\,$\mu$m of 0\farcs151 (Full Width at Half Maximum), the data in principle allow for the measurement of disk perturbations on time scales as short as 10 hours.  

\begin{figure*}[ht!]
\begin{interactive}{animation}{ratio_map_anim.mp4}
\includegraphics[width=18cm]{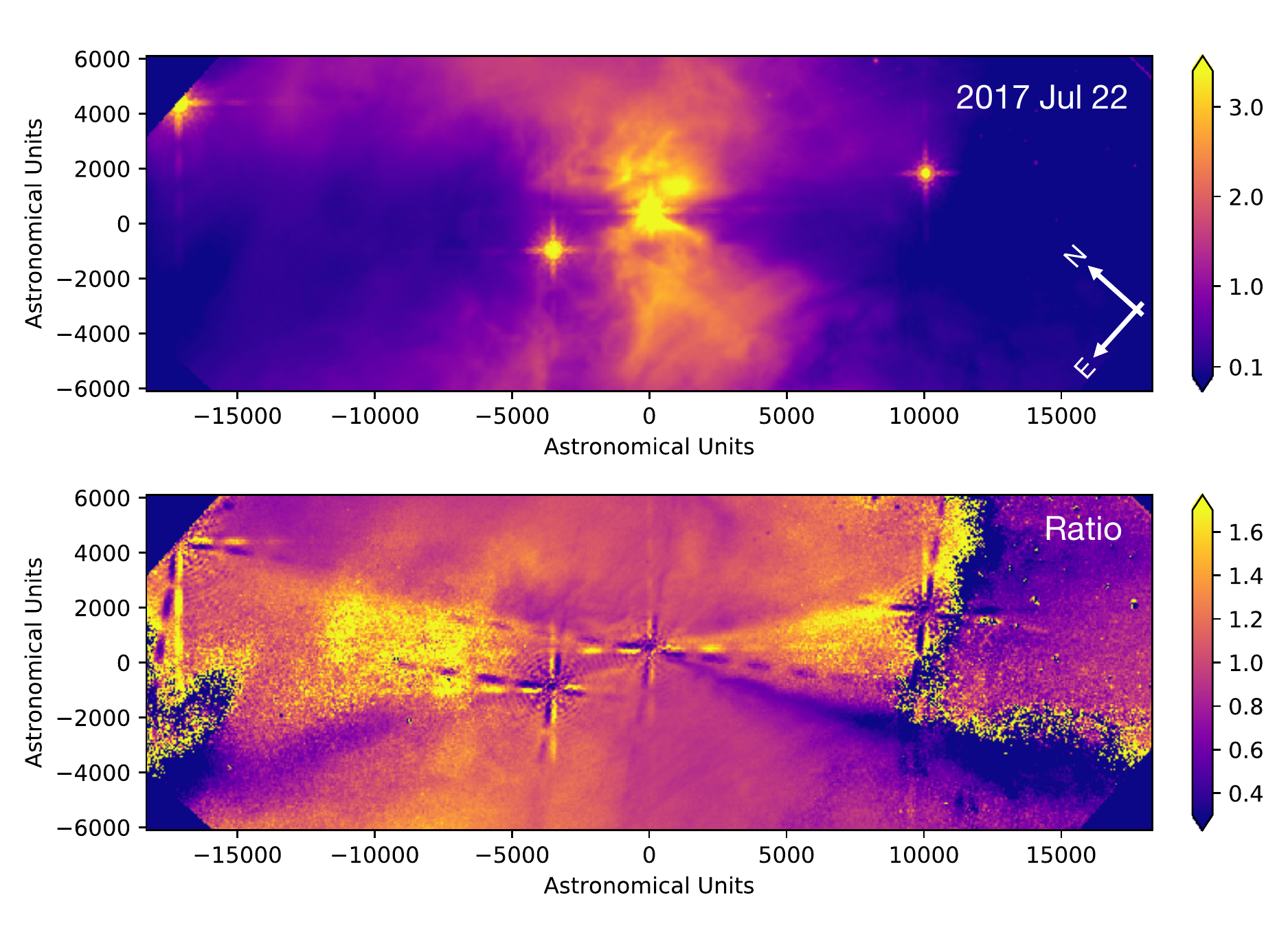}
\end{interactive}
\caption{Top panel: Rotated and scaled 1.6\,$\mu$m images of EC 82 for the two observing epochs, presented as an animation. The color scale is logarithmic and scaled to the same arbitrary unit to emphasize the shadow morphology. Bottom panel: The 1.6\,$\mu$m ratio image of the two epochs (Jul 2017/Aug 2018), with lighter colors indicating higher ratio. The resulting ratio image highlights the change in angle of the two shadow lobes. It also demonstrates that the change in angle happens in opposite directions for the two lobes. [This figure is animated in the published journal article.]
\label{fig:ratio_map}}
\end{figure*}

For the purpose of searching for variability, we compare the F160W (epoch 1) and F164N (epoch 2) filters, as these have overlapping band passes. While the F164N filter includes coverage of the [Fe II] line at 1.64\,$\mu$m, it is unlikely that any large scale difference in the shadow morphology is due to line emission, and we interpret the observed differences as being due to a change in broad wavelength illumination. 

Because the reflecting cloud contains significant substructure, changes in the shadow morphology are most apparent in differential or relative measurements between the two epochs. This is illustrated in Figure \ref{fig:ratio_map}, which shows the 1.6\,$\mu$m ratio image between the two epochs (August 2018/July 2017). The ratio image highlights the shadow as the dominant source of variability in the region. The most apparent change is that each lobe has changed its position angle across the full extent of the shadow by several degrees, and that the position angle change has occurred in opposite angular directions for the two lobes. That is, the western lobe moved in a clockwise direction, whereas the eastern lobe moved in a counter-clockwise direction between 2017 and 2018. The two shadow lobes are therefore not co-planar, with the 2017 epoch being more out-of-plane than the 2018 epoch. This is unequivocal evidence for non-axisymmetry of the system, although we cannot immediately distinguish between a non-axisymmetry of the disk itself, or of the illuminating source (see Section \ref{sec:discussion}).

\subsection{Retrieval of time-dependent disk parameters}

In this analysis, we quantify the time-dependence of disk structure by retrieving basic structural parameters as a function of time, based on equidistant cross sections of the shadow. We model the shadow using a simple pressure-supported and viscous disk model of the form \citep{Hartmann98}:

\begin{equation}
    \rho_{\rm d}(R,\theta) =  \\
    \frac{\Sigma_{\rm d}(R)}{\sqrt{2\pi} h_R R}  \times\exp{\Big[-\frac{1}{2} ((\pi/2-\theta)/h_R)^2\Big]}, 
\end{equation}

where $h_R=H/R$ is the disk scale height $H$ at radius $R$, in units of $R$, and $\theta$ is the polar angle measured from the axis of rotation. The scale height is parameterized as $h_R(R) = h_R(R_{\rm outer}) \times (R/R_{\rm outer})^{\psi}$. The dust surface density profile is also a power law, $\Sigma_d(R)\propto R^{\gamma}$. This formulation uses the approximation that the vertical coordinate $z= R\sin(\pi/2-\theta) \simeq R (\pi/2-\theta)$, appropriate for $\theta\simeq \pi/2$. The optical depth profile, $\tau$, is then a one-dimensional function of $\theta$:

\begin{figure*}[ht!]
\centering
\includegraphics[width=18cm]{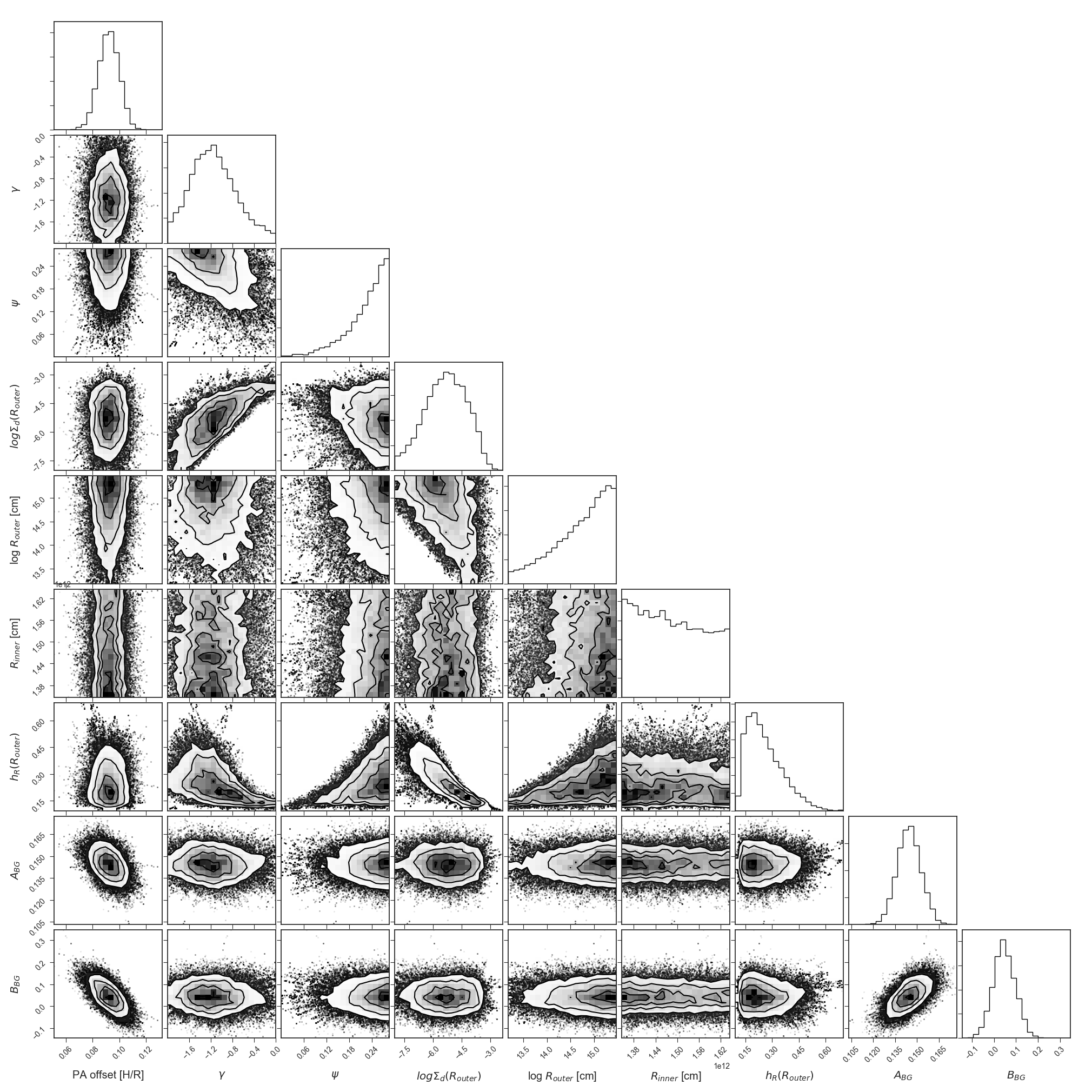}
\caption{Representative corner plot for the western lobe, epoch 2 at a distance of 7\farcs2 from EC 82, corresponding to 18.1 days. The corner plot demonstrates that the position angle offset (in units of H/R) is well-constrained, as is the background level, and the disk scale height, $(H/R)_0$. The radial surface density power law is consistent with an $R^{-1}$ dependence. Conversely, the inner and outer radii of the disk are not well-constrained.}
\label{fig:corner}
\end{figure*}

\begin{equation}
\tau(\theta) = \int_{R_{\rm inner}}^{R_{\rm outer}} \rho(R,\theta) C_{\rm ext} dR
\end{equation}

$C_{\rm ext}$ is the extinction coefficient at the observing wavelength, $\lambda$. This model assumes single scattering, but in practice the shadow is probably partly filled in either by multiple scattering, or by scattering of photons from other sources. Further, the distribution of scattering dust is not uniform. We therefore model the intensity profile of the shadow by adding a linear continuum, as well as a linear background component:

\begin{equation}
    I(\theta) = (I_C(\theta)-I_{\rm BG}) \times \exp({-\tau}) + I_{\rm BG},
\end{equation}

where $I_{\rm BG}=A_{\rm BG}+B_{\rm BG}\times \theta$. Finally, the intensity profile is convolved by a one-dimensional Gaussian kernel with a full width at half maximum of 0\farcs151 to simulate the WFC3 point spread function at 1.6~$\mu$m. We generally assume uninformative, or flat, priors for all parameters. The single exception is that we let the linear continuum be fixed, defined by the surface brightness on each side of the shadow, corresponding to a highly constrained prior. However, we do assume a flat prior for the background component.

We use {\tt emcee} \citep{Mackey13} to sample the posterior probability distribution of the model parameters. Because we wish to search for temporal variations in the shadow geometry within a single epoch, as well as between epochs, we retrieve independent probability distributions of slices of constant distance of the shadow from the source. To minimize the effect of any artifacts, we average each slice over 10 pixels in the radial direction. This translates into a time-resolution of $10\times 0\farcs08 = 0\farcs8 = 2.0$~days. 

For each slice, we use the following standard likelihood function:

\begin{equation}
    \ln(L)=-\frac{1}{2} (R^T C^{-1} R + \ln(\det(C)) + N_{\rm pix} \ln(2\pi)),
\end{equation}
where $R=\mathrm{Data - Model}$ is the residual and $C$ is the covariance matrix for the pixel errors.

\begin{figure}[ht!]
\centering
\includegraphics[width=8.5cm]{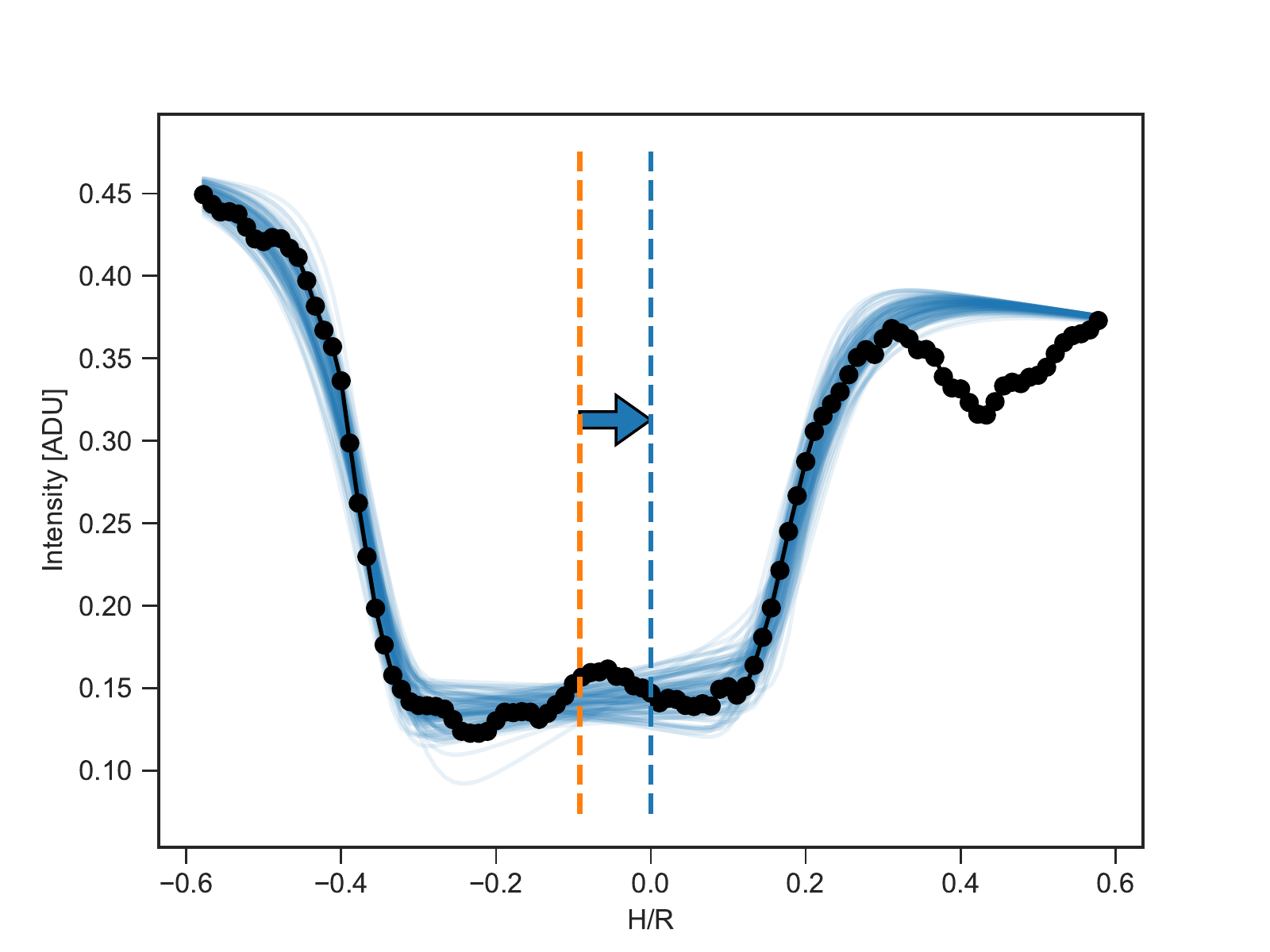}
\caption{Samples of converged models (blue lines) overlaid on the data (black line and points) for the same slice used for the corner plot in Figure \ref{fig:corner}. The arrow indicates the direction and magnitude of the retrieved angular offset of the shadow model, equivalent to a relative position angle.}
\label{fig:fit_quality}
\end{figure}

A representative corner plot for one of the slices is shown in Figure \ref{fig:corner} and an example fit is shown in Figure \ref{fig:fit_quality}. The probability distributions demonstrate that many parameters are degenerate, including the flaring index, as well as the inner and outer radii, but also that the disk scale height and position angle are well-determined. These are the two parameters that can be monitored over time. Finally, the background intensity is well-determined, although its slope correlates somewhat with the shadow position angle.

\subsection{Quantified shadow variability}
\label{sec:results}

\begin{figure*}[ht!]
\centering
\includegraphics[width=15cm]{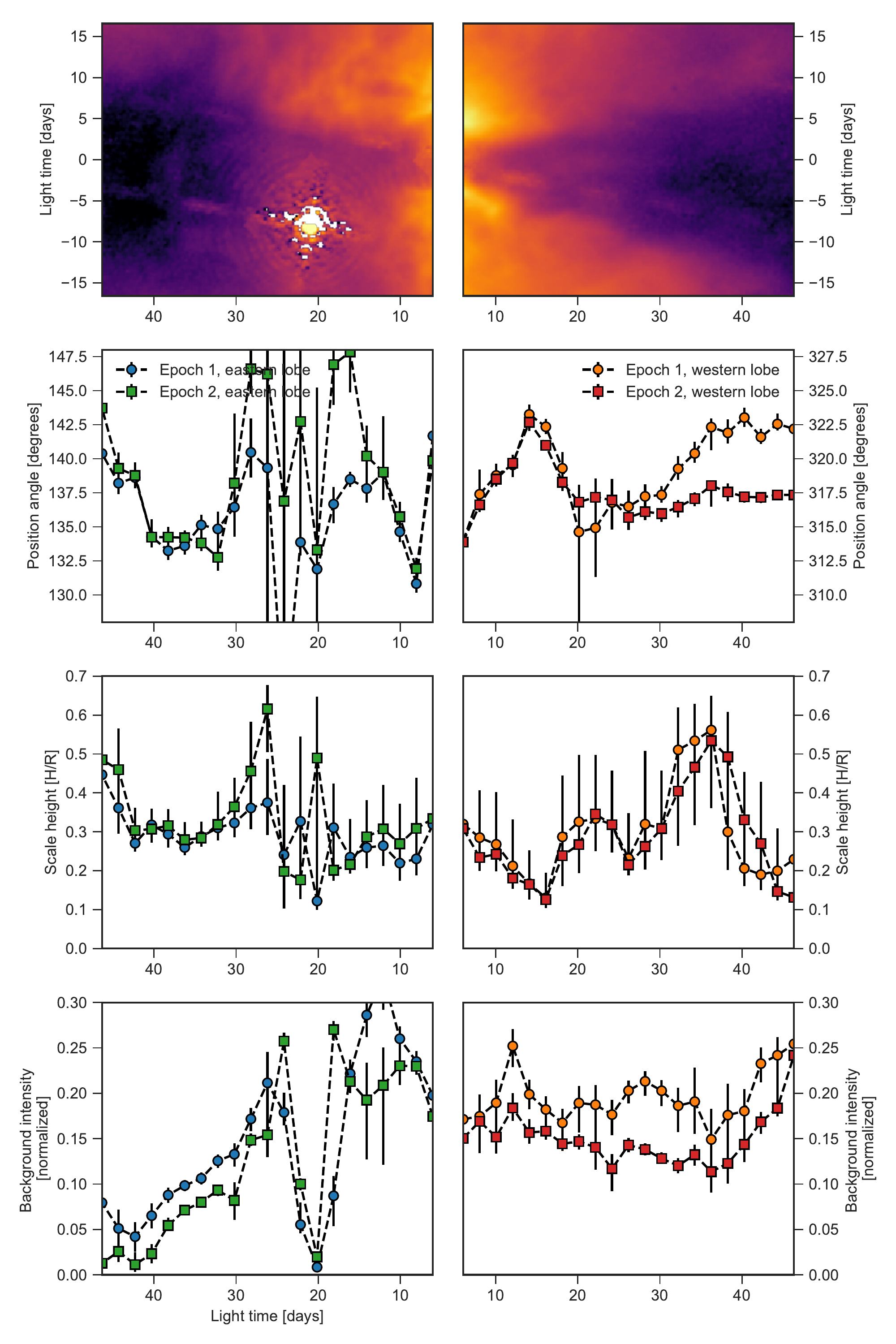}
\caption{Summary of constrained parameters as a function of light-travel time. The top panels show the epoch 2 image (F164N), while the lower three rows show the retrieved position angles, disk scale heights, and background intensities, respectively. Each epoch is indicated by a different color and symbol. The western lobe is much better defined, as the eastern lobe is strongly affected by the presence of a bright source (EC 86).}
\label{fig:alldata}
\end{figure*}

The subset of constrained parameters are summarized in Figure \ref{fig:alldata}. They include the position angle, the disk scale height, and the background intensity. The data for the two epochs are overlaid as a function of light-travel time relative to the time of observation. It is seen that there are features present in all parameters, and that these features reappear at the same relative light-travel time in both epochs. Because of this invariance, we interpret these features as being due to intrinsic physical structure in the scattering cloud, rather than properties of the shadow. For instance, there is a prominent offset in the position angle around 11-17 days in the western lobe, which is directly visible in the image as well. Because these physical structures have size scales of $\gtrsim 1000$~au, it is not expected that they will vary significantly over a 1-year time scale, as is indeed supported by their recurrence in both epochs. However, any {\it relative} difference between the two epochs, at a given distance, may be interpreted as being due to variations in the shadow illumination, in turn implying a corresponding variation in the disk structure. 

\begin{figure}[ht!]
\centering
\includegraphics[width=8.5cm]{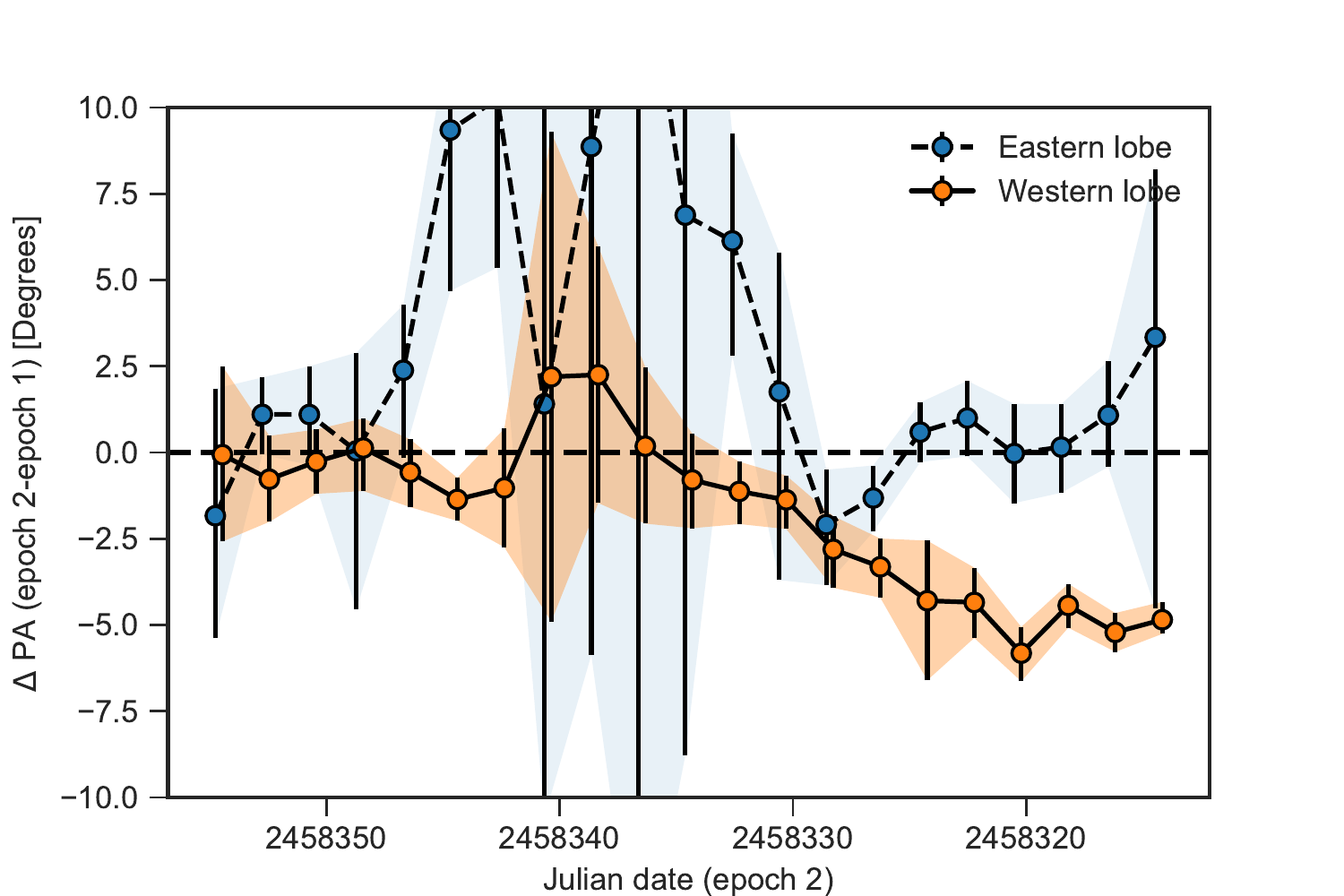}
\caption{Relative change in position angle between the two epochs as a function of Epoch 2 date. The values for each lobes are offset relative to each other by one day, for clarity.}
\label{fig:pa_relative}
\end{figure}

The strongest indicator of variability is the shadow position angle. In Figure \ref{fig:pa_relative}, the relative difference between the retrieved position angles is shown for both the eastern and western lobes (epoch 2 - epoch 1). Because we only have two epochs available, and the cloud features tend to be as strong as any shadow variability, we are not able to assign any relative difference to one epoch over the other. 

While the eastern lobe position angle is less certain due to the presence of a bright stellar source in the shadow, this metric recovers the trend seen in the ratio map in Figure \ref{fig:ratio_map} in which a clockwise shift in the western lobe (negative change in position angle) is matched by a counter-clockwise shift in the eastern lobe (positive change in position angle). Further, the western lobe displays a clear trend toward increasing negative angles as a function of time, moving the shadow from $\sim -1\deg$ to $\sim -5\deg$. Colloquially, one may think of the observed shadow variability as being reminiscent of the flapping of a bird's wings. 

There is not sufficient temporal coverage to detect any periodicity in the variation of the position angle. During the $\sim$45 days of detectable change available within a single image, the western lobe position angle changes in a manner consistent with a linear trend with time. That is, if there is periodicity in the variation, the period is at least $\sim 4\times 45\, = 180$\,days. Using Kepler's law, and assuming a stellar mass of 2.5\,$M_{\odot}$, this corresponds to a semi-major axis of $\gtrsim 0.85\,$AU. At the same time, a period is unlikely to be much greater than this, as an extrapolation of a sinusoidal signature would lead to position angle changes much larger than a few degrees, which is not seen in previous images of the Serpens shadow \citep[e.g.,][]{Gorlova10,Sugitani10}. Consequently, we find that the variability is consistent with orbital motion at $\gtrsim$1\,au and up to a few au, subject to confirmation by more extensive monitoring. 

Apart from the position angle, we can also look for changes in disk scale height. However, while Figure \ref{fig:alldata} shows variation in the retrieved scale height, the repeated pattern in the two epochs suggest that this variation is also dominated by the underlying cloud structure, rather than in shadow variability. Indeed, there is no significant indication of a systematic, relative difference in scale height between the two epochs, also not where the position angle difference is the greatest. The retrieved disk scale heights vary between $H/R=0.15-0.5$. The extreme ends of this range are probably dominated by cloud structure, but the preferred median of $H/R=0.25\pm 0.05$ can be interpreted as representative for the disk. While this disk scale height is much larger than that expected for a continuous flaring disk at 1\,au, it is roughly consistent with the height of a directly-irradiated puffed-up inner rim at the same radius \citep{Dullemond01}. Conversely, the shadow from a flared disk will be dominated by the outer disk scale height at large radii \citep[$\sim$20-100\,au, for a typical disk,][]{Hendler20}, but this is inconsistent with the short time scale of the variation. We therefore suggest that the shadowing material is dominated by a puffed-up inner disk, implying that the outer disk is self-shadowed, and that the short-term variability of the large shadow is evidence of self-shadowing within the disk itself. 

\subsection{Evolutionary stage of EC 82}
\label{sec:evolution}
The properties and evolutionary stage of EC 82 are uncertain, in part due to the fact that its edge-on disk obscures the central star, decreases its apparent luminosity, and changes the shape of the spectral energy distribution of the star-disk system. While it has long been thought that the central star is a solar-mass object, the recently revised distance to Serpens of $\sim 436$\,pc \citep{OrtizLeon17} increases the luminosity of EC 82 to 30\,$L_{\odot}$, making it likely that EC 82 is actually an intermediate-mass star.

\begin{figure}[ht!]
\centering
\includegraphics[width=8.5cm]{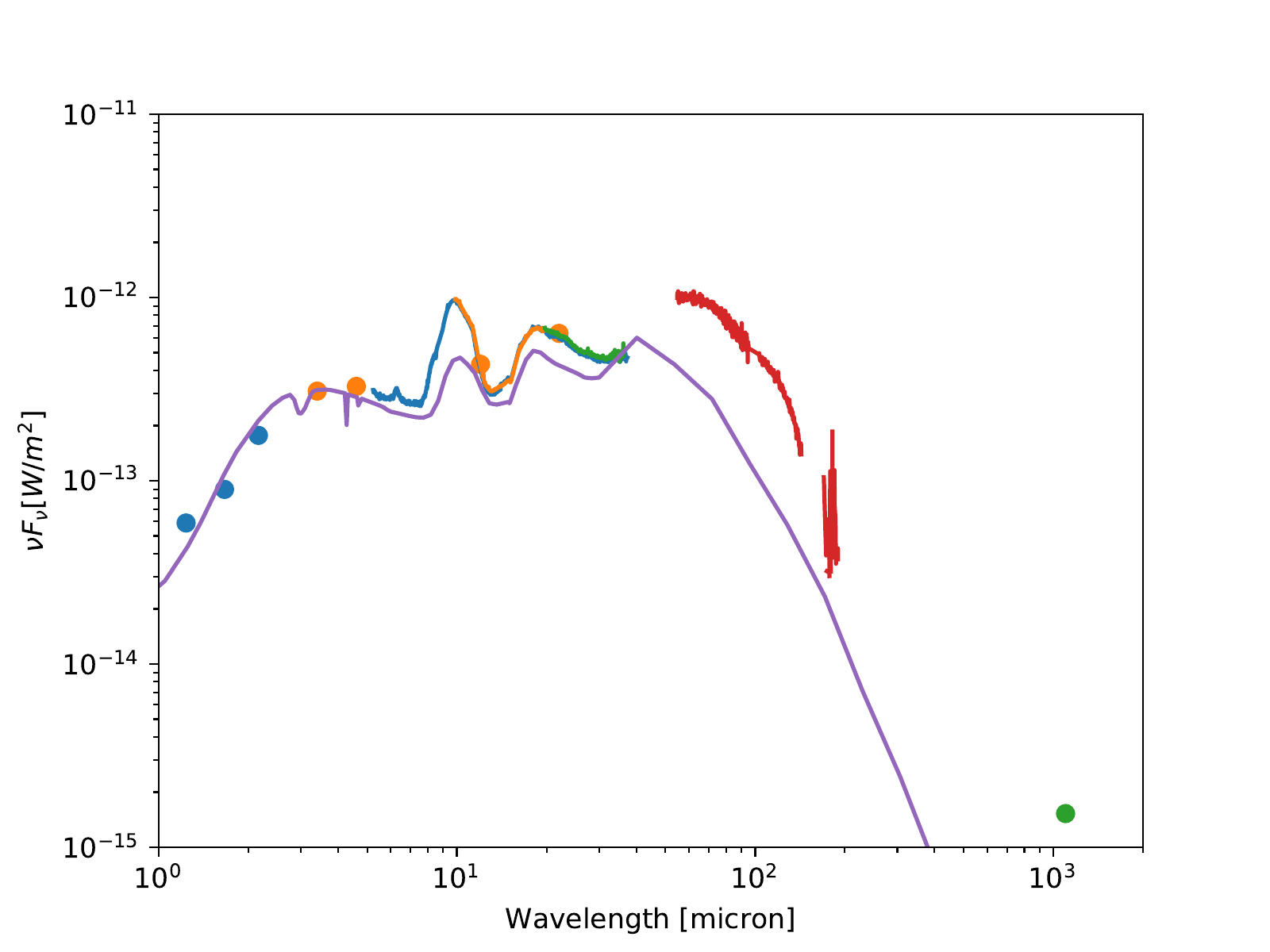}
\caption{Observed Spectral Energy Distribution of EC 82. The data include photometry from 2MASS, WISE, \cite{Dunham15}, and spectroscopy from Spitzer \citep{Pontoppidan10,Lebouteiller11} and Herschel \citep{Green16}. The solid curve shows the model from \cite{Pontoppidan05}, derived before the PACS spectroscopy was available.}
\label{fig:sed}
\end{figure}

The spectral energy distribution (SED) of EC 82 is shown in Figure \ref{fig:sed}. The 2-24\,$\mu$m logarithmic spectral index defined as:

\begin{equation}
 \alpha = \frac{d\log\lambda F_{\lambda}}{d\log\lambda} = 0.3
\end{equation}
formally identifies the source as a class I object \citep{Dunham15}. However, this identification is confounded by the edge-on geometry, which suppresses the short-wavelength range of the SED. The presence of strong 10 and 20\,$\mu$m silicate emission features are difficult to reconcile with the presence of any substantial protostellar envelope, which would invariably create deep silicate and ice absorption features due to the presence of significant column densities of cold dust toward the central infrared source \citep{Robitaille07}. Consequently, if EC 82 were viewed at a more face-on angle, it is likely to display an SED typical for a protoplanetary disk. In the context of the Herbig stars, it is difficult to predict if the face-on SED would be typical of Group I (flared) or II (self-shadowed) disks \citep{Meeus01, Acke09}. Since the shadow is likely to be formed at small radii, based on the short time scale for the shadow variability, EC 82 disk is likely self-shadowed at large radii. 

Is EC 82 a classical, full disk, or a more evolved, transitional disk? The system appears to have significant amounts of hot material (up to 1500\,K) near the central star, as evidenced by strong veiling at 1--2\,$\mu$m \citep{Doppmann05, Gorlova10}, strong mid-infrared silicate emission, and strong rovibrational CO emission lines \citep{Banzatti15}. The mid-infrared slope of the SED is flat \citep[see Figure 5 and][]{Dunham15}, indicative of a full disk. \citet{Pontoppidan05} found that it was difficult to model the strong silicate emission features while considering the edge-on orientation and the presence of a large surrounding reflection nebula. Their solution was to model the system using a very low-mass disk ($M_{\rm disk}\sim 10^{-5}\,M_{\odot}$), indicative of a highly evolved system similar to a debris disk, and a flat density distribution of the surrounding cloud. Figure \ref{fig:sed} shows the spectral energy distribution of EC 82, including the Herschel PACS spectra superimposed on the best-fitting model from \citet{Pontoppidan05}. Finally, this model found that the reflection nebula corresponds to a near-infrared extinction of $A_J=1.4\,$mag along the line of sight, estimated both by the SED fit, as well as the observed depth of the 3.08\,$\mu$m water ice absorption band. In this work, we do not attempt to re-fit the radiative transfer model, but note that the new PACS spectra and millimeter photometry point to a more massive outer disk than that proposed in \citet{Pontoppidan05}, making it more likely that EC 82 is a classical protoplanetary disk. However, adding a more massive outer disk while still fitting the strong emission features from the inner disk may require the use of a flat, self-shadowed outer disk component. 

\section{Discussion}
\label{sec:discussion}
The hydrodynamics of protoplanetary disks govern both star- and planet-formation, but are difficult to constrain by direct observation. Even with high-resolution imaging, typically only a snapshot in time is available. Traditional tracers of disk dynamics include instantaneous velocity tracers using high-resolution line spectroscopy \citep{Hughes11, Teague19}, or indirect accretion tracers, such as hydrogen recombination lines \citep{Muzerolle98,Salyk13}. In this paper, we have shown that giant disk shadows may be used to provide a continuous measure of disk motions with $<1$ day resolution using a combination of relatively infrequent (every 40-50 days) imaging and the finite light-travel time across the shadow. A drawback is that giant disk shadows, such as that in Serpens, are rare, and are thus only available for a very small number of disks. 

Of particular interest is the potential for strong periodicity in the shadow position angle, as this may indicate an origin in interactions with a low-mass companion, including planetary-mass objects. If the variability is due to a forced perturbation, e.g., from a low-mass companion or planet, strong periodicity is likely. However, the anti-symmetry of the position angle change (the ``flapping'') in the two lobes suggests a {\it quadrupolar} disk warp rather than a bipolar warp, which would produce a symmetric ``wobble''. It is not presently clear if there is a theoretical basis for such a quadrupolar warp, and further modeling is required to test this idea.

\subsection{An alternate explanation for the variability}
\begin{figure}[ht!]
\centering
\includegraphics[width=8.5cm]{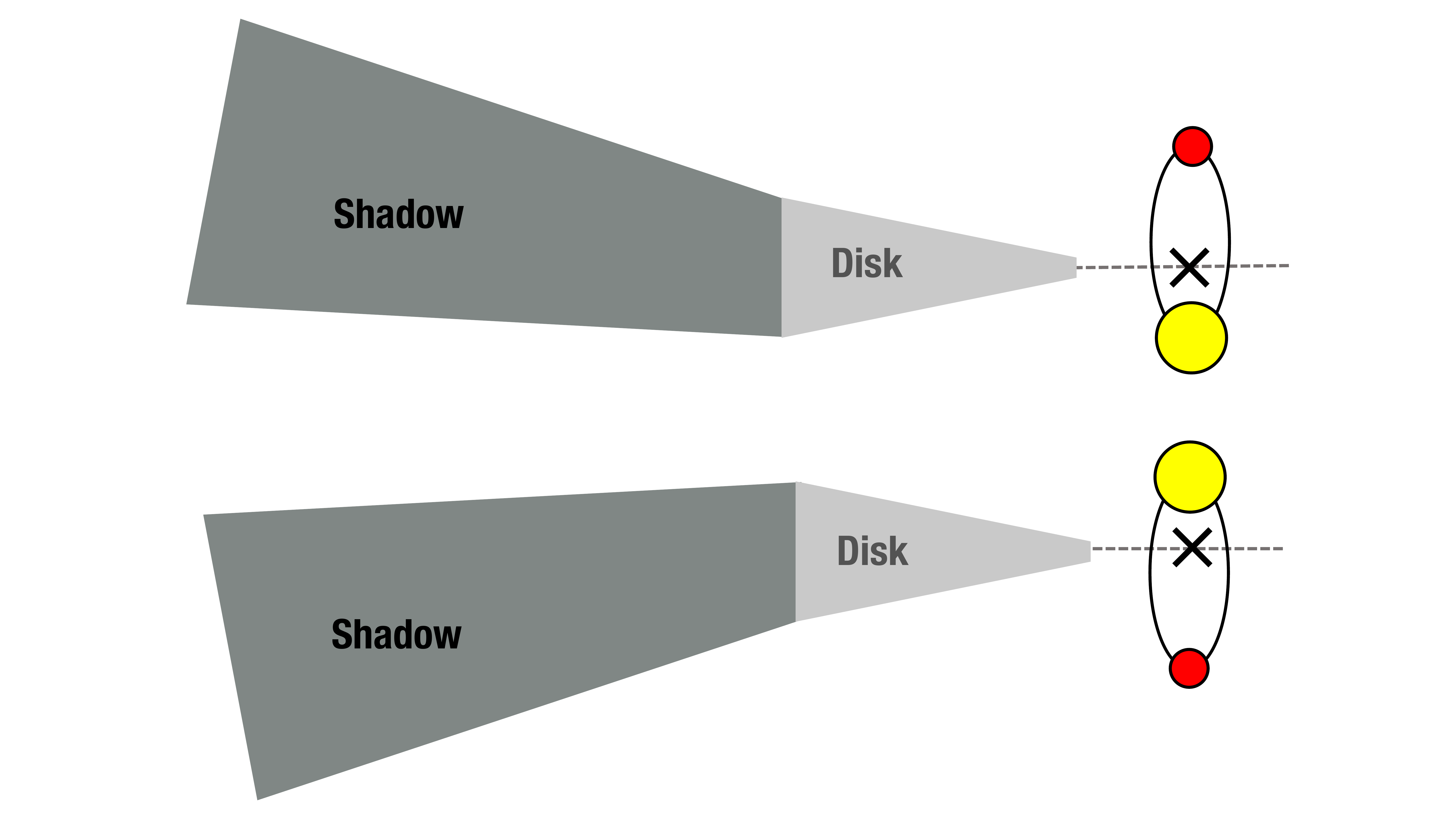}
\caption{Sketch showing an alternative explanation for the shadow variability, in which an unequal-mass binary orbiting out of the disk plane shifts the photo-center of the illumination. This predicts that the inner disk is cleared-out of material and that the position angle variation is strongly periodic. The sketch is not to scale.} 
\label{fig:binary_cartoon}
\end{figure}

We have discussed a model in which the variability is caused by an orbiting, or precessing, quadrupolar disk warp. However, there is an alternative explanation: A low-mass companion to EC 82, orbiting out of the plane of the disk, may move the photo-center of the source relative to the disk plane. This would also lead to non-planar (flapping) change in observed shadow position angle. This scenario is illustrated in Figure \ref{fig:binary_cartoon}. It requires an unequal-mass (or unequal-luminosity) binary, as an equal-mass binary would not move the photo-center at any point in its orbit. The main argument against a stellar binary may be that this scenario is at odds with the presence of an optically thick inner disk, as evidenced by the spectral energy distribution of the system (see Section \ref{sec:evolution}). That is, a low-mass stellar companion on a $\gtrsim$ 160 day orbit will likely clear the inner several au of any material, leaving a deficit in the mid-infrared parts of the SED \citep{Cieza10}. Further, the dynamic action of such a binary on the disk itself is likely significant, so this does not exclude that part of the variability is still due to a disk warp, in this case induced by a stellar binary. There is to our knowledge no current evidence for a spectroscopic binary in EC 82, although the available stellar spectroscopy is probably not constraining due to the presence of strong veiling \citep{Doppmann05}. This scenario predicts consistent and strong periodicity of the shadow position angle, and can therefore be excluded if future monitoring fails to detect such a periodic signature. 

\subsection{Other indicators of variability in EC82}
CO fundamental rovibrational lines (at 4.7\,$\mu$m) have been observed with VLT/CRIRES (R$\sim$100,000), and show three distinct components: a very broad emission component from the disk, a broad absorption component from an envelope or outer disk, and a narrow, extended emission component that may arise in an outflow \citep{Brown13}.  Two epochs of observations with Keck/NIRSPEC (R$\sim$25,000), in April 2002 and July 2004, show some differences in both the emission and absorption components (C. Salyk, priv. comm.).  Both cyclical and non-cyclical explanations are consistent with these observations.  However, a small source Doppler shift in July results in poor telluric correction near the line center for the 2004 data, so further spectroscopic monitoring of the infrared CO lines is needed to confirm this result.

\section{Conclusion}
We have found that the giant disk shadow projected by the young star EC 82 in the Serpens core is variable. In particular, the position angle of the shadowing material relative to the stellar photocenter changes by several degrees over time scales of a year. The large angular size of the shadow corresponds to a light travel time of more than 45 days, allowing for detailed constraints on the dynamics of the EC 82 disk using the magnification effect of the long shadow. Because the variability time scale is relatively short, we conclude that the occulting material is located within a few au of the central star, and that any disk component at larger radii is likely self-shadowed. We suggest that further monitoring of the disk shadow from a stable platform such as Hubble, or the upcoming James Webb Space Telescope, offers a unique opportunity to constrain, in real time, the hydrodynamics of terrestrial planet-forming regions. Based on just two epochs of imaging, we cannot determine if the variability is periodic, and further imaging on a months to years cadence is required to establish, or reject, periodicity and a potential connection to a low-mass perturber.

\acknowledgments
This work is based on observations made with the NASA/ESA Hubble Space Telescope, obtained from the data archive at the Space Telescope Science Institute. STScI is operated by the Association of Universities for Research in Astronomy, Inc. under NASA contract NAS 5-26555. This work is based in part on observations made with the Spitzer Space Telescope, which was operated by the Jet Propulsion Laboratory, California Institute of Technology under a contract with NASA. Herschel is an ESA space observatory with science instruments provided by European-led Principal Investigator consortia and with important participation from NASA. KMP and TP are supported by a NASA ROSES XRP grant NNX17AB60G S005.

\vspace{1.5cm}
%

\vspace{5mm}

\facilities{HST(WFC3), WISE, 2MASS, Herschel(PACS), Spitzer(IRS)}


\software{This paper made use of the {\tt astropy} package \citep{astropy13}, {\tt matplotlib} \citep{Hunter07}, the {\tt emcee} package by \cite{Mackey13}, and {\tt DrizzlePac} \citep{2012ascl.soft12011S}.}

\bibliographystyle{aasjournal}



\end{document}